\documentclass[aps,onecolumn,11pt,floatfix,altaffilletter,preprintnumbers,tightenlines,showpacs,showkeys,notitlepage]{revtex4-1}

\usepackage[colorlinks=true,citecolor=blue,linkcolor=blue]{hyperref}
\usepackage[normalem]{ulem}
\usepackage{amsmath,amssymb}
\usepackage{epsfig}
\usepackage{epsfig}
\pdfoutput=1
\usepackage{graphicx}               % Standard graphics package
\usepackage{url}
\usepackage{color}
\usepackage{multirow}
\usepackage{placeins}
\usepackage[dvipsnames]{xcolor}
\usepackage{epstopdf}

\usepackage{tikz}
\usetikzlibrary{trees}
\usetikzlibrary{decorations.pathmorphing}
\usetikzlibrary{decorations.markings}

\definecolor{jblue}  {RGB}{20,50,100}
\definecolor{npurple}  {RGB} {153, 51, 204}
\definecolor{wred}   {RGB}{217,0,56}
\definecolor{white}   {RGB}{255,255,255}

\definecolor{korange}   {RGB}{235, 80,  43}
\definecolor{korange2}   {RGB}{245, 100,  63}
\definecolor{kyelloworange}   {RGB}{255, 210,  110}
\definecolor{kyelloworange2}   {RGB}{240, 170,  90}
\definecolor{kred}   {RGB}{204,  102, 153}
\definecolor{kpurple}   {RGB}{153,  61, 190}
\definecolor{kpurplelight}   {RGB}{213,  161, 230}

\usepackage{cleveref}
\usepackage{subfigure} % for subfigures
\usepackage{lipsum}
\definecolor{red}{rgb}{1.0, 0, 0}
 \usepackage{gensymb}
\allowdisplaybreaks

\bibpunct{[}{]}{,}{n}{}{,}

\newcommand{\parenbar}[1]{\overset{
            \raisebox{-0.15em}{\scalebox{.4}{\textbf{(}}}
            \raisebox{-0.3em}{{\hspace{.03em}--\hspace{.05em}}}
            \raisebox{-0.15em}{\scalebox{.4}{\textbf{)}}}} {#1}}

\def\thesection{\Roman{section}}

\pacs{}

\begin{document}

\preprint{MITP/16-116}

%=============================================================================
\title{Sterile Neutrinos and Flavor Ratios in IceCube}

\author{Vedran Brdar}   \email{vbrdar@uni-mainz.de}
\author{Joachim Kopp}   \email{jkopp@uni-mainz.de}
\author{Xiao-Ping Wang} \email{xiaowang@uni-mainz.de}
\affiliation{PRISMA Cluster of Excellence and
             Mainz Institute for Theoretical Physics,
             Johannes Gutenberg-Universit\"{a}t Mainz, 55099 Mainz, Germany}
%=============================================================================

\begin{abstract}
  The flavor composition of astrophysical neutrinos observed in neutrino
  telescopes is a powerful discriminator between different astrophysical
  neutrino production mechanisms and can also teach us about the particle
  physics properties of neutrinos. In this paper, we investigate how the
  possible existence of light sterile neutrinos can affect these flavor ratios.
  We consider two scenarios: (i) neutrino production in conventional
  astrophysical sources, followed by partial oscillation into sterile states;
  (ii) neutrinos from dark matter decay with a primary flavor composition
  enhanced in tau neutrinos or sterile neutrinos. Throughout the paper, we
  constrain the sterile neutrino mixing parameters from a full global fit to
  short and long baseline data. We present our results in the form of flavor
  triangles and, for scenario (ii), as exclusion limits on the dark matter mass
  and lifetime, derived from a fit to IceCube high energy starting
  events and through-going muons. We argue that identifying a possible flux
  of neutrinos from dark matter decay may require analyzing the flavor
  composition as a function of neutrino energy.
\end{abstract}

\maketitle

%-----------------------------------------------------------------------------
\section{Introduction}
\label{sec:intro}
%-----------------------------------------------------------------------------

In the era of multi-messenger astronomy, it is fair to say that we can not only
see the Universe (using telescopes that cover the entire electromagnetic
spectrum), but also hear the Universe (using gravitational wave
detectors~\cite{Abbott:2016blz}), and taste the Universe (using neutrino
telescopes).  Regarding the latter, we are of course referring to the
possibility of measuring the flavor of high-energy neutrinos and of using this
information to learn about the properties of astrophysical neutrino sources and
about the properties of neutrinos themselves.
Indeed, the non-trivial information offered by the flavor ratios of astrophysical
neutrino fluxes has been studied ever since the discovery of neutrino oscillations
\cite{Barenboim:2003jm,     % Flavor triangles, neutrino mixing, CPT violation, \nu decay
  Cirelli:2004cz,           % Short section on impact of sterile neutrinos
  Xing:2006uk,              % Parameterization in terms of two angles \xi and \zeta
  Lipari:2007su,            % Spectra and flavor ratios (+ uncertainties) as of 2007
  Pakvasa:2007dc,           % Expansion in U_{e3}, 45-\th_{23}; non-standard initial ratios
  Blennow:2009rp,           % Impact of NSI in production/detection on flavor ratios
  Esmaili:2009dz,           % Measuring \theta_{13}, \delta_{CP} in flavor ratios
  Lai:2009ke,               % Sensitivity study for distinguishing different sources
  Choubey:2009jq,           % Parameterization 1:n:0.
  Bhattacharya:2010xj,      % Lorentz/CPT violation, decay, decoherence, ...
Hollander:2013im,
Keranen:2003xd},          % Flavor ratios with sterile neutrinos
and the topic has received a tremendous further boost in 2013
\cite{Rajpoot:2013dha,      % Impact of sterile neutrinos
  Mena:2014sja,             % The 2014 analysis by Mena, Palomarez-Ruis, Vincent
  Xu:2014via,               % Constraints on flavor ratios from unitarity of U_{PMNS}
  Fu:2014isa,               % Propagation matrix P; several aspects of measuing flavor ratios
  Palomares-Ruiz:2015mka,   % Comprehensive analysis including spectrum + flavor
  Palladino:2015zua,        % Theorists' analysis of flavor ratios, E>60 TeV, 3 yrs of data
  Aartsen:2015ivb,          % The IceCube paper on flavor ratios
  Kawanaka:2015qza,         % Impact of \pi/\mu acceleration before decay
  Palladino:2015vna,        % Parameterization of vacuum oscillation: P_0, P_1, P_2
  Arguelles:2015dca,        % General parameterization of new physics
  Bustamante:2015waa,       % Flavor ratios for arbitrary initial composition
  Aartsen:2015ita,          % Combination of 6 IceCube analyses
  Gonzalez-Garcia:2016gpq,  % Impact of NSI in Earth matter
Li:2016kra,                 % "echo method" for improving flavor identification
Shoemaker:2015qul,
Chatterjee:2013tza,
Chen:2014gxa,
Nunokawa:2016pop},              
thanks to the discovery of a high energy astrophysical neutrino flux by the IceCube
collaboration
\cite{Aartsen:2013jdh,      % The Science paper on the discovery of astrophysical \nu
  Aartsen:2014gkd,          % UHE neutrinos in 3 years of IceCube data
  Aartsen:2015ivb,          % The IceCube paper on flavor ratios
Aartsen:2015ita}.           % Combination of 6 IceCube analyses

In this paper, we discuss in particular how the flavor ratios of ultra-high
energy astrophysical neutrinos are affected in the presence of sterile
neutrinos that have sizeable mixing with active neutrinos.  Motivation for
such scenarios comes for instance from the long-standing anomalies observed by
some short-baseline oscillation experiments, in particular
LSND~\cite{Aguilar:2001ty}, MiniBooNE~\cite{Aguilar-Arevalo:2013pmq}, SAGE and
GALEX~\cite{Acero:2007su, Giunti:2010zu}, as well as several reactor neutrino
experiments~\cite{Mueller:2011nm, Mention:2011rk, Huber:2011wv} (see also
\cite{Hayes:2013wra, Fang:2015cma, Hayes:2016qnu}).
Even though these results appear to be in some tension with non-observations
in other experiments~\cite{Kopp:2011qd, Conrad:2012qt, Archidiacono:2013xxa,
Kopp:2013vaa, Mirizzi:2013kva, Giunti:2013aea, Gariazzo:2013gua,
Collin:2016rao}, they have motivated a multifarious experimental
program aimed at testing them~\cite{Abazajian:2012ys}.
Compared to previous analyses studying high-energy astrophysical neutrino
fluxes and flavor ratios in the presence of sterile
states~\cite{Cirelli:2004cz, Rajpoot:2013dha, Hollander:2013im,
Hasenkamp:2016pme, Cherry:2016jol}, we will comprehensively include
constraints from short baseline oscillation experiments. We
will achieve this by using the numerical fitting codes underlying the global
fit from ref.~\cite{Kopp:2013vaa}. Moreover, we will include the possibility
that the initial flux of high-energy neutrinos is dominated by tau neutrinos or
sterile neutrinos in some part of the energy spectrum. This can happen for instance in
scenarios where PeV-scale dark matter (DM) particles decay to high energy active
or sterile neutrinos.
The latter scenario is particularly interesting because the observed
flavor ratios would depend mainly on the active--sterile mixing angles.

The outline of the paper is as follows: in \cref{sec:methods} we will collect
the relevant analytic formulas for computing flavor ratios of high-energy
neutrinos, and we will describe how we implement the constraints from
oscillation experiments in our analysis. In \cref{sec:astro+DM-sources}, we
will then present our results for the case of astrophysical
neutrino sources and for a primary flux dominated by tau neutrinos or
sterile neutrinos. We will
show the accessible regions in the ``flavor triangle'' --- the well-known
ternary plot illustrating the fractions of electron, muon and tau neutrinos
relative to the total flux of active neutrinos.  In \cref{sec:astro-fit} we
will discuss two toy models for heavy DM particles decaying to active or
sterile neutrinos.  These models serve as illustrative examples for scenarios
with non-standard initial flavor composition. We will also constrain the parameter
space of our toy models by fitting the HESE (high energy starting event) and
TGM (through-going muon) data from IceCube.  In \cref{sec:summary}, we summarize and conclude.

%-----------------------------------------------------------------------------
\section{Computing Flavor Ratios in the Presence of Sterile Neutrinos}
\label{sec:methods}
%-----------------------------------------------------------------------------

In the following, we discuss the production, propagation, and detection of high
energy astrophysical neutrinos in the presence of an additional eV scale sterile neutrino
$\nu_s$ that is a singlet under the $\text{SU(3)}_{\text{c}} \times
\text{SU(2)}_{\text{L}} \times \text{U(1)}_{\text{Y}}$ gauge group of the
Standard Model (SM).
High energy astrophysical neutrinos are usually assumed to be produced
in the decays of high energy pions, which in turn originate in collisions
of high energy cosmic protons with other nucleons or with photons.
The ensuing flavor composition of the
primary flux is then $(\Phi_{\nu_e} : \Phi_{\nu_\mu} : \Phi_{\nu_\tau} :
\Phi_{\nu_s}) \sim (1 : 2 : 0 : 0)$, unless muons rapidly lose energy before
decaying, in which case the flavor composition of high energy neutrinos changes
to $(\Phi_{\nu_e} : \Phi_{\nu_\mu} : \Phi_{\nu_\tau} : \Phi_{\nu_s}) \sim (0 :
1 : 0 : 0)$~\cite{Kashti:2005qa, Winter:2014pya,Waxman:1997ti,Rachen:1998fd}. Here, $\Phi_{\nu_\alpha}$ is the initial flux of neutrinos of flavor $\alpha$.  Alternative scenarios include neutrino
production in neutron decay with initial flavor composition $(1 : 0 : 0 :
0)$~\cite{Lipari:2007su}, or in decays of charm mesons, leading to a flavor
composition of $(1 : 1 : 0 : 0)$~\cite{Choubey:2009jq}. In the literature, an
initial flavor ratio of $(0:0:1:0)$ is also studied for
completeness~\cite{Arguelles:2015dca}, despite the lack of a plausible
astrophysical production scenario. In this work we will study a simple model for PeV dark matter decaying dominantly to the third lepton generation. We will also consider the possibility
that part of the astrophysical neutrino flux is initially sterile, i.e.\ has
flavor composition $(0 : 0 : 0 : 1)$. Such a flux could originate, for
instance, from dark matter decay into sterile neutrinos or from a dark sector
with complex dynamics of its own, admitting the existence of ``dark
astrophysical accelerators'' \cite{Blinnikov:2009nn,Foot:2014mia}.

Neutrinos are initially created as flavor eigenstates, i.e.\ coherent superpositions
of mass eigenstates, that begin to oscillate as they propagate through space.
However, as the distance from a typical astrophysical neutrino
source is much larger than the coherence length, only the averaged
effect of these oscillations is observable at Earth. Oscillation probabilities
can thus be computed by treating the initial flux as an incoherent
superposition of mass eigenstates. This approach is further justified by the
fact that different neutrino mass eigenstates propagate at slightly different group
velocities. By the time a neutrino arrives at Earth, its different mass eigenstate
components are therefore separated in space and time and can no longer be
detected coherently. With these considerations in mind, the flavor
conversion probabilities for high energy neutrinos in a world with
$n$ (active\,+\,sterile) neutrino flavors are~\cite{Giunti:2007ry}
\begin{align}
  P_{\nu_\alpha \to \nu_\beta}
    = P_{\nu_\beta \to \nu_\alpha}
    = \delta_{\alpha\beta}
    - 2\sum_{k>j} \text{Re} \big[ U_{\alpha k}^* U_{\beta k}
                                  U_{\alpha j} U_{\beta j}^* \big]
    = \sum_{j=1}^n |U_{\alpha j}|^2 |U_{\beta j}|^2 \,,
  \label{eq:Pab}
\end{align}
where $U$ is the $n \times n$ leptonic mixing matrix. In this work we focus on
the case $n=4$ and  adopt the parameterization \cite{deGouvea:2008nm, Kopp:2013vaa} 
\begin{align}
  U = R^{34}(\theta_{34}) \tilde{R}^{24}(\theta_{24},\delta_2)
      R^{23}(\theta_{23}) R^{14}(\theta_{14})
      \tilde{R}^{13}(\theta_{13},\delta_0) \tilde{R}^{12}(\theta_{12},\delta_1) \,,
  \label{eq:UPMNS}
\end{align}
where $R^{ij}(\theta_{ij})$ is a rotation matrix in the $ij$ plane and
$\tilde{R}^{ij}(\theta_{ij},\delta_k)$ is a rotation matrix supplemented with
an additional phase factor $\delta_k$. Each of these matrices is unitary, which
implies that their product $U$ is unitary as well. In the 3+1 model (three active neutrinos
and one sterile neutrino) considered here, there are six mixing angles $\theta_{ij}$
and three physical phases. (Majorana phases are omitted as they are not observable in
oscillation experiments.)

In general, we have the following 12 parameters to consider 
\begin{equation}
 \Theta \equiv (\theta_{12},
\theta_{13}, \theta_{23}, \delta_{0}, \theta_{14}, \theta_{24}, \theta_{34},
\delta_1, \delta_2, \Delta m_{21}^2, \Delta m_{31}^2, \Delta m_{41}^2). 
\end{equation}
As $\Delta m_{21}^2, \Delta m_{31}^2$ are well measured and not related to a rotation matrix, we fix them at 
$|\Delta m_{21}^2|=7.5\times10^{-5} \rm{eV}^2$ and $|\Delta m_{31}^2|=2.4\times10^{-3}\rm{eV}^2$.
 In order to explore the viable parameter space for neutrino oscillations, we randomly generate
$10^7$ parameter sets $\Theta$.

For each parameter set, we randomly draw $\Delta m_{41}^2$  between 
$0.1~ \rm{eV}^2$ and $10 ~\rm{eV}^2$. As for the mixing angles and
phases, we take them to be distributed according to the Haar measure
\cite{Haba:2000be,Haar,Brdar:2015jwo}, which reads in the four flavor case
\begin{equation}
  d\Theta = d(\sin^2 \theta_{12}) \,
            d(\sin^2 \theta_{23}) \,
            d(\cos^4 \theta_{13}) \,
            d(\cos^6 \theta_{14}) \,
            d(\cos^4 \theta_{24}) \,
            d(\sin^2 \theta_{34}) \,
            d\delta\,d\delta_1\,d\delta_2 \,.
  \label{eq:Haar}
\end{equation}
In other words, the distributions of $\sin^2 \theta_{12}$, $\sin^2 \theta_{23}$,
$\cos^4 \theta_{13}$, etc., are flat.
By generating parameter points according to the Haar measure, we ensure
that our prior distribution is independent of the chosen parameterization
of the mixing matrix~\cite{Haba:2000be}.
Already when generating parameter points, we restrict the three standard mixing
angles $\theta_{12}$, $\theta_{23}$, $\theta_{13}$ to their experimentally
allowed $3\sigma$ ranges based on a three-flavor fit using the global fitting
code from ref.~\cite{Kopp:2013vaa}.
Afterward, we process all generated points
with the same code, but including sterile neutrinos. Our code includes the
experimental data sets listed in \cref{tab:exps}.  This allows us to assign a
global $\chi^2$ value to each parameter point.

\begin{table}
  \renewcommand{\arraystretch}{1.3}
  \begin{center}
    \begin{minipage}{12cm}
    \begin{ruledtabular}
    \begin{tabular}{lc}
      Experiment                       & Oscillation channel(s) \\
      \hline
      Short and long-baseline reactors & $\bar{\nu}_e\to\bar{\nu}_e$ \\
      KAMLAND                          & $\bar{\nu}_e\to\bar{\nu}_e$ \\
      Gallium                          & $\nu_e\to\nu_e$ \\
      Solar neutrinos                  & $\nu_e\to\nu_e$,
                                          neutral current (NC) data  \\
      LSND/KARMEN $^{12}$C             & $\nu_e\to\nu_e$ \\
      CDHS                             & $\nu_\mu\to\nu_\mu$  \\
      MiniBooNE                        & $\parenbar{\nu_{\mu}} \to \parenbar{\nu_{e}}$,
                                         $\parenbar{\nu_{\mu}} \to \parenbar{\nu_{\mu}}$ \\
      MINOS                            & $\nu_\mu\to \nu_\mu$, NC data \\
      Atmospheric neutrinos            & $\parenbar{\nu_{\mu}} \to \parenbar{\nu_{\mu}}$ \\
      LSND                             & $\bar{\nu}_\mu\to \bar{\nu}_e$ \\
      KARMEN                           & $\bar{\nu}_\mu\to \bar{\nu}_e$ \\
      NOMAD                            & $\nu_\mu\to \nu_e$ \\
      E776                             & $\parenbar{\nu_{\mu}} \to \parenbar{\nu_{e}}$ \\
      Icarus                           & $\nu_\mu\to \nu_e$ \\
    \end{tabular}
    \end{ruledtabular}
    \end{minipage}
  \end{center}
  \caption{Oscillation experiments used in our analysis, based on the global fit from
    ref.~\cite{Kopp:2013vaa}. The column labeled ``Oscillation channel(s)''
    indicates whether a given experiment is measuring disappearance or appearance
    of (anti)neutrinos.}
  \label{tab:exps}
\end{table}  

We then use Bayesian
statistics to determine the credible intervals for the parameters (see
for instance ref.~\cite{kendall1994vol}).
For each parameter set $\Theta$, the probability of obtaining the observed
data is given by
\begin{align}
  P(\text{data} | \Theta)
    = \exp\bigg[-\frac{\chi^2(\Theta)}{2} \bigg].
  \label{bayes1}
\end{align}
The unconditional probability $P(\text{data}) =
\int\!d\Theta\,P(\Theta)\,P(\text{data}|\Theta)$ in our case is obtained by
integrating $P(\text{data}|\Theta)$ over all parameter sets. 
We choose a flat prior, $P(\Theta) = \text{const.}$, in accordance with our assumption
that parameter sets have a flat distribution in the Haar measure.
We now apply Bayes' theorem \cite{kendall1994vol}
\begin{align}
  P(\Theta|\text{data}) = \frac{P(\text{data}|\Theta) \, P(\Theta)}{P(\text{data})}
  \label{eq:bayes-theorem}
\end{align}
to obtain the posterior probability $P(\Theta|\text{data})$ of a given
parameter set $\Theta$, given the data.
The $\alpha\%$ credible region in parameter space is
then obtained by ordering the parameter points in descending order in the
posterior probability: $P(\Theta_1|\text{data}) > P(\Theta_2|\text{data}) >
P(\Theta_3|\text{data}) > \dots$. We then choose an $i_\text{max}$ such that
\begin{align}
  \sum_{i=1}^{i_\text{max}} P(\Theta_i|\text{data}) = \frac{\alpha}{100} \,.
\end{align}
All parameter points $\Theta_i$ with $i < i_\text{max}$ are included in the
$\alpha\%$ credible interval, all other points are excluded.

%-----------------------------------------------------------------------------
\section{Predicted Neutrino Flavor Ratios}
\label{sec:astro+DM-sources}
%-----------------------------------------------------------------------------

\begin{figure}
  \centering
  \begin{tabular}{cc}
    \includegraphics[width=0.48\textwidth]{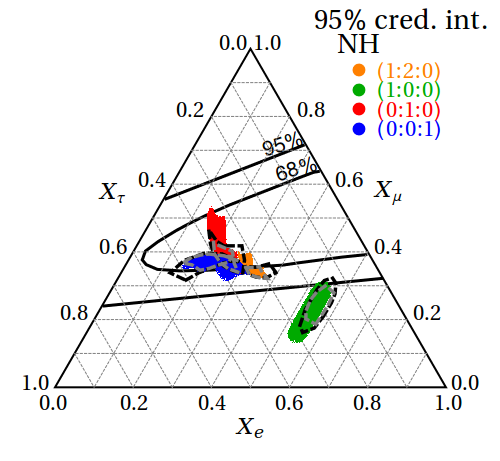} &
    \includegraphics[width=0.42\textwidth]{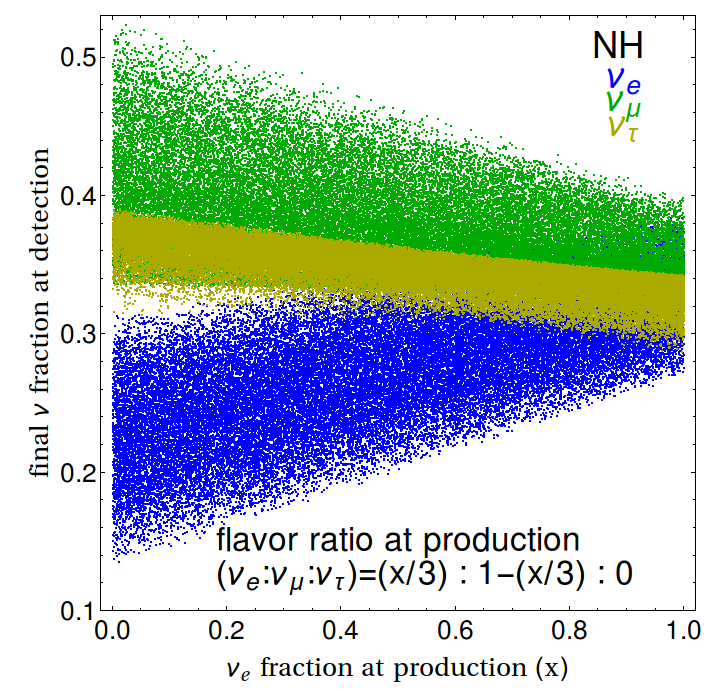} \\
    (a) & (b)
  \end{tabular}
  \caption{(a) Flavor composition of high-energy astrophysical neutrinos at Earth
    for different assumptions on the initial flavor composition
    $\Phi_{\nu_e}^{\text{in}} : \Phi_{\nu_\mu}^{\text{in}} :
    \Phi_{\nu_\tau}^{\text{in}}$, assuming only standard three-flavor oscillations.
    The size of the colored regions represents the uncertainty in the three-flavor
    oscillation parameters based on the global fit developed in
    ref.~\cite{Kopp:2013vaa}. For comparison, we also show results based on the
    more recent fits from ref.~\cite{Gonzalez-Garcia:2014bfa} (black dashed contours)
    and ref.~\cite{Esteban:2016qun} (gray dashed contours).
    Black solid contours indicate
    the flavor ratios preferred by IceCube data at the 68\% and 95\% confidence
    level \cite{Aartsen:2015knd}. Note that we show only results for normal
    neutrino mass ordering (NH) since the plot for inverted ordering would be
    almost identical.  (b) Variation of the fractional $\nu_e$, $\nu_\mu$ and
    $\nu_\tau$ fluxes at Earth as a function of the initial $\nu_e$ fraction $x$,
    assuming an initial flavor composition of the form $(\frac{x}{3} :
    1-\frac{x}{3} : 0)$.}
  \label{fig:triangle-sm}
\end{figure}

A useful tool in studying neutrino flavor ratios is a ternary diagram (``flavor
triangle'') as shown in \cref{fig:triangle-sm} (a).  The three axes correspond
to the fraction of neutrinos detected as $\nu_e$, $\nu_\mu$, and $\nu_\tau$,
respectively.  The predicted flavor ratios at Earth in the absence of sterile
neutrinos are shown as colored regions in \cref{fig:triangle-sm} for
different assumptions on the composition of the primary flux.
These regions correspond to 95 \% credible interval. In obtaining
these credibility regions, we have followed the approach described
in \cref{sec:methods}, based on the global fit from ref.~\cite{Kopp:2013vaa},
but we show for comparison also the preferred regions based
on the more recent results from ref.~\cite{Gonzalez-Garcia:2014bfa} (dashed black
contours) and ref.~\cite{Esteban:2016qun} (dashed gray contours).
To obtain the latter, we assume the likelihood distributions
for $\sin^2\theta_{12}$, $\sin^2\theta_{13}$, and $\sin^2\theta_{23}$
to be Gaussian, with central values and widths taken from
ref.~\cite{Gonzalez-Garcia:2014bfa,Esteban:2016qun}.
 For the phase $\delta_0$,
we assume a flat distribution in the interval $[-\pi, \pi]$.
Note that we show only results assuming a normal mass hierarchy (NH) here.
We have checked that the plot changes very little for the inverted hierarchy case.
We also note good agreement between our results and those of
refs.~\cite{Arguelles:2015dca,Bustamante:2015waa}.
Comparing the colored region in \cref{fig:triangle-sm} (a) to the 68\%
and 95\% CL exclusion regions provided by IceCube (black contours)~\cite{Aartsen:2015knd},
we observe that only a neutron decay source with initial flavor composition
$(1:0:0)$ is in some tension with the data.

In \cref{fig:triangle-sm} (b), we illustrate in more detail how the flavor
composition of astrophysical neutrinos at Earth in the standard 3-flavor scenario
varies with the initial composition, which we assume here
to be of the form $(\frac{x}{3} : 1-\frac{x}{3} : 0)$. Varying $x$ between $0$ and $1$
thus interpolates between the $(0:1:0)$ and $(1:2:0)$ flavor compositions.
The observable dispersion in the flavor composition is again due to uncertainties in
the measured standard mixing parameters.

\begin{figure}
  \centering
  \includegraphics[width=\columnwidth]{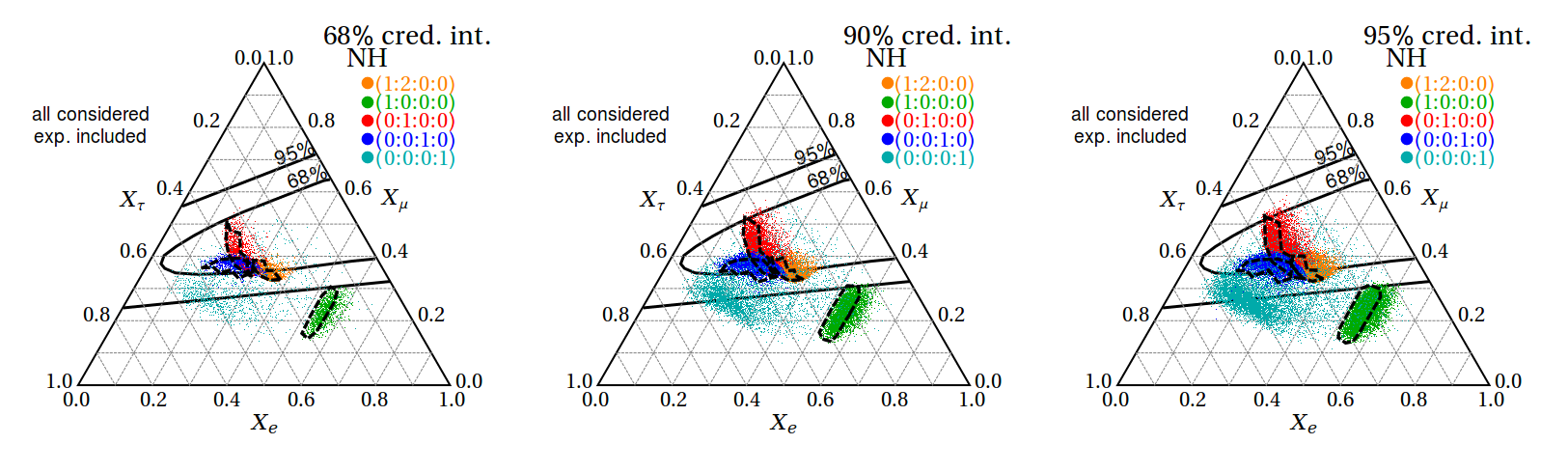}
  \caption{Flavor composition of high-energy astrophysical neutrinos at Earth in
    the presence of sterile neutrinos.  For the initial flavor composition, we
    consider in addition to the scenarios from \cref{fig:triangle-sm} also the
    possibility of a purely sterile initial flux, $(\Phi_{\nu_e} :
    \Phi_{\nu_\mu} : \Phi_{\nu_\tau} : \Phi_{\nu_s}) = (0:0:0:1)$. We show in
    color the parameter points corresponding to the 68\% (left), 90\% (middle)
    and 95\% (right) credibility intervals from a global fit to short and long
    baseline data \cite{Kopp:2013vaa} (see text for details).  The regions
    delineated by dashed black lines are the corresponding intervals without
    sterile neutrinos from \cref{fig:triangle-sm}.  Large solid black contours
    correspond to the IceCube constraint on the flavor ratios
    \cite{Aartsen:2015knd}.  Once again, the results shown here are for normal
    neutrino mass ordering. 
  }
  \label{fig:triangle-nu-s-all}
\end{figure}

Coming to the impact of sterile neutrinos on the flavor composition of
astrophysical neutrinos, we show in \cref{fig:triangle-nu-s-all} the reachable
parts of the flavor triangle including oscillations into sterile neutrinos. The
oscillation parameters are constrained to lie in the 68\% (left), 90\%
(middle), or 95\% (right) credible interval obtained from the Bayesian global
fit described in \cref{sec:methods}. By comparing to the credible intervals in
the 3-flavor case (black dashed contours in \cref{fig:triangle-nu-s-all}), we
see that, for initial flavor compositions consisting only of active neutrinos,
the situation is very similar to the standard scenario without sterile
neutrinos.  Large deviations from the 3-flavor situation are not possible given
that active--sterile mixing angles are constrained to be $\lesssim
\mathcal{O}(10\%)$~\cite{Kopp:2013vaa}.  On the other hand, for flux components
that are initially purely sterile (cyan regions in
\cref{fig:triangle-nu-s-all}), the observed flavor ratios at Earth are
preferably in the lower left part of the flavor triangle, with relatively
large $\nu_\tau$ component and much smaller $\nu_e$ and $\nu_\mu$ admixtures.
This is also easily understandable: $\theta_{34}$ is much more weakly
constrained than $\theta_{14}$ and $\theta_{24}$, so that $P_{\nu_s \to
\nu_\tau}$ can be large~\cite{Kopp:2013vaa}.  Comparing to IceCube constraints
(black contours in \cref{fig:triangle-nu-s-all}), we see that an initial
neutrino flux consisting purely of $\nu_e$ (e.g.\ from neutron decay) is
disfavored, as in the standard 3-flavor case. A purely sterile initial flux is
still allowed, depending on the exact values of the mixing angles. Note,
however, that IceCube constraints are based on the whole neutrino energy
spectrum above $\text{few} \times 10$\,TeV, and that an initial flux consisting
only of $\nu_s$ throughout this energy range seems not very plausible
theoretically. 

\begin{figure}
  \centering
  \includegraphics[width=\columnwidth]{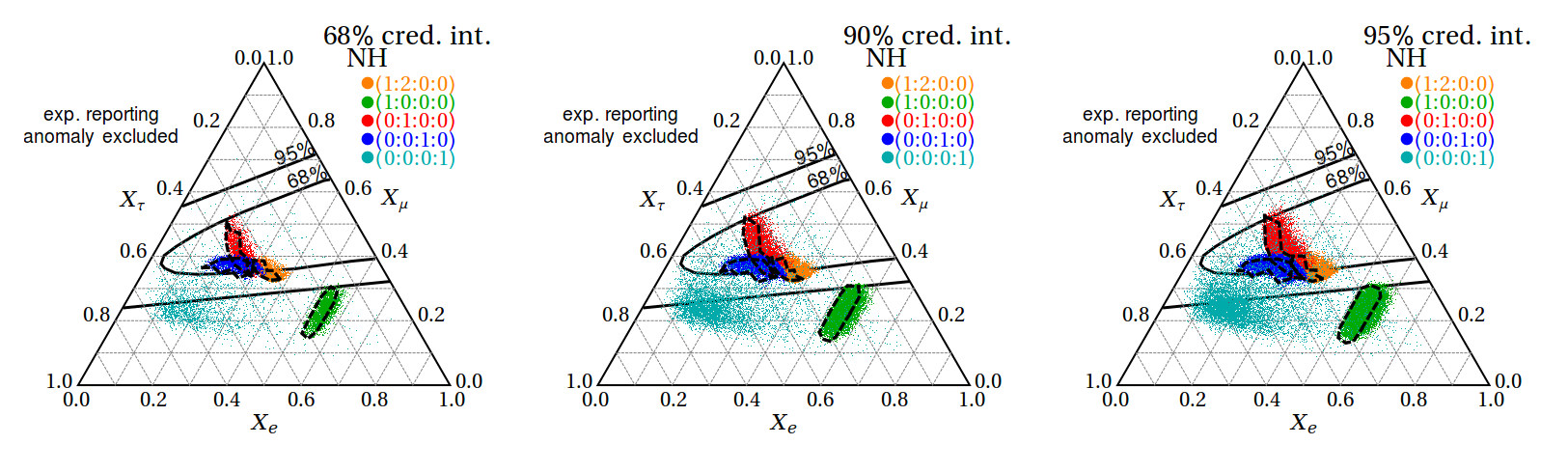}
  \caption{Same as \cref{fig:triangle-nu-s-all}, but excluding the anomalous data
    sets from LSND, MiniBooNE, short-baseline reactor experiments, and Gallium
    experiments in constraining the sterile neutrino mixing parameters.
    }
  \label{fig:triangle-nu-s-no-anomalies}
\end{figure}

For illustration, we have also investigated how the results shown in
\cref{fig:triangle-nu-s-all} change when the short-baseline anomalies
from LSND, MiniBooNE, short-baseline reactor experiments, and Gallium
experiments are disregarded, see \cref{fig:triangle-nu-s-no-anomalies}.
We see virtually no change compared to \cref{fig:triangle-nu-s-all},
which indicates that the global fit is dominated by the null searches.
The largest differences are observed for purely sterile initial flux, where
the preferred flavor ratios at Earth are shifted towards pure $\nu_\tau$ flavor
when the anomalies are disregarded. The reason is that limits
on $\theta_{14}$ and $\theta_{24}$ are more stringent in this case, while
constraints on $\theta_{34}$ are not affected by the anomalous data sets.

\begin{figure}
  \centering
  \begin{tabular}{cc}
    \includegraphics[width=0.42\textwidth]{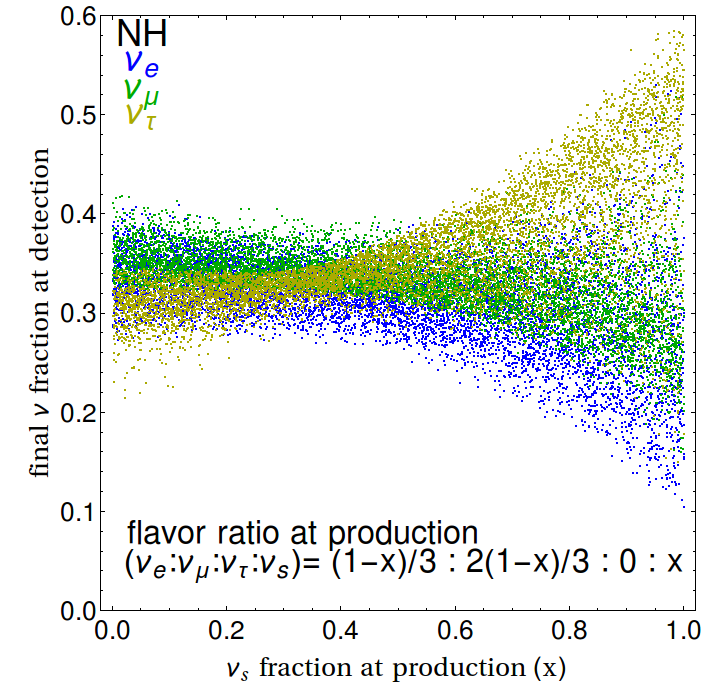} &
    \includegraphics[width=0.42\textwidth]{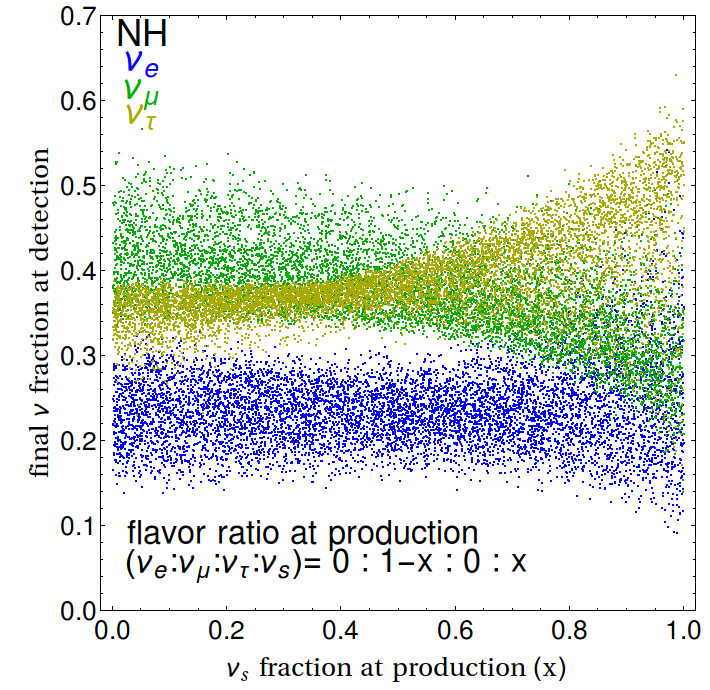} \\
    (a) & (b)
  \end{tabular}
  \caption{The individual flavor fractions at Earth for varying
    initial $\nu_s$ admixture. In panel (a), we take the initial flavor
    ratios to be of the form $\Phi_{\nu_e} : \Phi_{\nu_\mu} : \Phi_{\nu_\tau}
    : \Phi_{\nu_s} = (\frac{1-x}{3}:\frac{2(1-x)}{3}:0:x)$, with $0 < x < 1$. An initial
    flavor composition of this form could arise, for instance, from
    an astrophysical flux from pion decay plus a $\nu_s$ flux from
    dark matter decay. In panel
    (b) we use the parameterization $(0:1-x:0:x)$ with $0 < x < 1$ for
    the initial flux, corresponding for instance to a muon-damped astrophysical
    source plus $\nu_s$ from DM decay.
    The parameter points shown in both panels correspond to the 95\% credible
  interval.}
  \label{fig:varying-nu-s}
\end{figure}

Even though we have considered scenarios with initial flavor composition
$(0:0:0:1)$ in \cref{fig:triangle-nu-s-all,fig:triangle-nu-s-no-anomalies}, we
have emphasized that a purely sterile initial flux is not very likely, neither
from a theoretical point of view nor from looking at the IceCube data.
Therefore, we illustrate in \cref{fig:varying-nu-s} how the predicted flavor
fractions at Earth change as a function of the $\nu_s$ admixture to the initial
flux. The two parameterizations for the initial flavor composition shown in
this figure, $(\frac{1-x}{3}:\frac{2(1-x)}{3}:0:x)$ and $(0:1-x:0:x)$
correspond to an admixture of sterile neutrinos (e.g.\ from DM decay) to an
astrophysical flux from pion decay, or from pion decay with strong muon energy
loss (muon-damped source), respectively.

%-----------------------------------------------------------------------------
\section{Dark matter decay and fits to IceCube data}
\label{sec:astro-fit}
%-----------------------------------------------------------------------------

To illustrate how flavor ratios of high-energy neutrinos can be used to
probe and discriminate between particle physics models, we discuss in the
following two toy scenarios in which neutrinos are produced not only by
conventional astrophysical sources, but also in the decay of PeV-scale DM
particles. The two toy models of interest to us are:
\begin{itemize}
  \item {\bf Model~1: Heavy right-handed neutrino DM.} \\
    In this scenario, the DM particles $N$ are fermionic and are total singlets under
    the SM gauge group. They are allowed to decay to SM particles via the
    neutrino portal interactions
    \begin{align}
      \mathcal{L}_{\text{int}}
        \supset  \sum_\alpha  y_{\alpha} \bar{N} \, \tilde{H}^\dag\, L_\alpha \,+\, \text{h.c.} \,.
      \label{eq:L-NR-model}
    \end{align}
    Here, $H$ is the SM Higgs boson, $\tilde{H} = i \sigma^2 H^*$, and
    $L_\alpha$ is the SM lepton doublet of flavor $\alpha$. We assume that the
    couplings of $N$ to first and second generation leptons are suppressed
    ($y_\tau \gg y_e,\,y_\mu$).  With  this assumption, the model predicts that
    in half of the $N$ decays, a monoenergetic tau neutrino with initial
    spectrum $dN_\nu/dE_\nu = \delta(m_\text{DM}/2 - E_\nu)$ is produced. Here,
    $m_\text{DM}$ is the mass of the DM particle $N$. Note that the model also
    predicts a secondary neutrino flux at lower energies, coming from decays of
    the Higgs boson produced in association with the monoenergetic neutrinos,
    the decay products of the $W$ bosons produced together with charged
    leptons, and from the decays of $Z$ bosons.  We compute all secondary
    neutrino fluxes using ref.~\cite{Cirelli:2010xx} (with electroweak
    corrections based on ref.~\cite{Ciafaloni:2010ti}) and cross-checked with
    those presented in \cite{Higaki:2014dwa,Esmaili:2014rma}.
        
    We require that $y_\tau$ is tiny,
    so that the lifetime of $N$~\cite{Higaki:2014dwa},
    \begin{align}
      \tau_\text{DM} = \frac{4\pi}{\, m_\text{DM} \, y_\tau^2}
    \end{align}
    is much larger than the age of the Universe and that gamma ray constraints are
    satisfied~\cite{Esmaili:2014rma}. The $N$ mass is taken
    to be $\mathcal{O}(\text{PeV})$.
    The smallness of $y_\tau$ could for instance be explained in scenarios with
    warped extra dimensions, where the wave function overlap between SM fields
    living on the infrared brane and $N$ living on the ultraviolet brane can be
    minuscule. 
   
  \item {\bf Model~2: Scalar DM decaying to sterile neutrinos.}
    This model is based on the Yukawa interaction
    \begin{align}
      \mathcal{L}_{\text{int}} \supset y_{2} \, \phi \bar\nu_s \nu_s \,+\, \text{h.c.} \,,
    \end{align}
    where $\phi$ is the scalar DM particle with mass $m_\text{DM} \sim
    \mathcal{O}(\text{PeV})$ and $\nu_s$ is eV-scale sterile neutrino
    introduced in \cref{sec:methods}. We again assume the Yukawa coupling $y_2$
    to be tiny so that the lifetime of $\phi$,
    \begin{align}
      \tau_\text{DM} = \frac{4\pi}{m_\text{DM} \, y_2^2}
    \end{align}
    is much larger than the age of the
    Universe.  The sterile neutrinos $\nu_s$ are assumed to mix with the active ones
    to lead to observable signals in IceCube.
\end{itemize} 
In computing the expected neutrino fluxes at the detector, we need to distinguish
between neutrinos from DM decay in the Milky Way and neutrino from DM decay in
distant galaxies.  The galactic component of the flux is not affected by redshift,
while the energy spectrum of extragalactic neutrinos is smeared out towards
lower energies because of contributions from high-redshift galaxies.  The differential
flux of galactic neutrinos from DM decay can be written as~\cite{Bai:2013nga,
Esmaili:2012us,Murase:2012xs,Dev:2016qbd,Fong:2014bsa,Cohen:2016uyg}
\begin{align}
  \frac{dJ_\text{gal}}{dE_\nu}
    \approx 1.7 \times 10^{-5} \, \text{cm}^{-2} \, \text{s}^{-1} \, \text{sr}^{-1} \,
            \bigg( \frac{1\,\text{GeV}}{m_\text{DM}} \bigg)
            \bigg( \frac{10^{26}\,\text{s}}{\tau_\text{DM}} \bigg)
            \times \frac{dN_\nu(E_\nu)}{dE_\nu} \,,
          \label{eq:gal-flux}
\end{align}
where $dN_\nu(E_\nu) / dE_\nu$ is the initial neutrino spectrum at production.
For neutrinos from extragalactic DM decay, we have instead~\cite{Esmaili:2012us}
\begin{align}
  \frac{dJ_\text{extragal}}{dE_\nu}
    &= \frac{\Omega_\text{DM} \rho_c}{4 \pi m_\text{DM} \tau_\text{DM}}
      \int_0^\infty \!dz\, \frac{1}{H(z)} \frac{dN_\nu\big((1+z)E_\nu\big)}{dE_\nu}
                                                                           \\
    &\approx 1.4 \times 10^{-5} \, \text{cm}^{-2} \, \text{s}^{-1} \, \text{sr}^{-1}
            \bigg( \frac{1\,\text{GeV}}{m_\text{DM}} \bigg)
            \bigg( \frac{10^{26}\,\text{s}}{\tau_\text{DM}} \bigg)
                                                \nonumber\\
    &\hspace{3cm} \times
            \int_0^{\frac{m_\text{DM}}{2 E_\nu}-1} \!dz\,
              \frac{1}{\sqrt{\Omega_\Lambda+\Omega_m(1+z)^3}}
            \frac{dN_\nu\big((1+z)E_\nu\big)}{dE_\nu} \,,
  \label{eq:J-extragal}
\end{align}
where $\Omega_\text{DM}$, $\Omega_m$, and $\Omega_\Lambda$ are the dark matter density,
total matter density, and dark energy density of the Universe, respectively,
all expressed in units of the critical density $\rho_c$. Moreover, $z$ is the redshift,
and $E_\nu$ is the neutrino energy at Earth, after redshifting.
Note that we have assumed the Universe to be perfectly transparent to neutrinos,
which is usually a safe assumption \cite{Esmaili:2012us}. 

We will fit four-year high energy starting events (HESE) data
from IceCube \cite{Aartsen:2015zva, Aartsen:2014gkd} as well as two years
of through-going muons (TGM) data \cite{Aartsen:2015rwa}. Note that we do not make use of
the more recent 6-year TGM analysis \cite{Aartsen:2016xlq} which, unlike
the 2-year one~\cite{Aartsen:2015rwa}, is not accompanied by a full digital data
release.  In particular, the response tensor $A_\text{eff}(E_\mu,E_\nu)$,
which relates the true neutrino
energy $E_\nu$ to the reconstructed muon energy $E_\mu$, is not available. For HESE data,
we make the simplification of equating the true neutrino energy and the
electromagnetic energy deposited in the detector.

The total number of DM-induced HESE events expected in IceCube in an energy interval $[E_a,
E_b]$ is obtained as
\begin{align}
  N_{\text{DM}}(E_a < E_\nu < E_b)
    = \int_{E_a}^{E_b} \!dE_\nu\,  \,\sum_{f=e,\bar{e},\mu,\bar\mu,\tau,\bar\tau} \Delta\Omega\,\Delta t \, A^f_\text{eff}(E_\nu) \,
        \bigg(\frac{dJ^{f,\text{osc}}_\text{extragal}}{dE_\nu}
            + \frac{dJ^{f,\text{osc}}_\text{gal}}{dE_\nu}\bigg) \,,
  \label{eq:N-DM}
\end{align}
where $\Delta t$ is the length of the data taking period, $A_\text{eff}(E_\nu)$ is
the energy dependent effective area of the detector \cite{Aartsen:2013jdh}, and $\Delta\Omega$ is the
solid angle range to which the experiment is sensitive. For high energy starting
events, we have $\Delta\Omega = 4\pi$. The superscript ``osc'' in
$dJ^\text{osc}_\text{extragal}/dE_\nu$ and $dJ^\text{osc}_\text{gal}/dE_\nu$
denotes that neutrino oscillations are taken into account in these fluxes
(as opposed to \cref{eq:gal-flux,eq:J-extragal}, which give unoscillated
fluxes).  For model~1, we use the best fit oscillation parameter values from
\citep{Gonzalez-Garcia:2014bfa}, while for model~2 we use the best fit parameter
set from our sterile neutrino fits, based on \cite{Kopp:2013vaa}.

Regarding TGM, the corresponding event number in $i$-th reconstructed muon
energy bin  can be expressed as 
\begin{align}
  N(E_a < E_\mu < E_b)
    = \int_{E_a}^{E_b} \! dE_\mu \int \! dE_\nu
        \sum_{f=\mu,\bar\mu,\tau,\bar\tau} \,
        \Delta\Omega \, \Delta t \,
        \frac{dA_\text{eff}^f(E_\mu, E_\nu)}{dE_\mu}
        \bigg(\frac{dJ^{f,\text{osc}}_\text{extragal}}{dE_\nu}
            + \frac{dJ^{f,\text{osc}}_\text{gal}}{dE_\nu}\bigg) \,.
  \label{eq:N-DM-TGM}
\end{align}
Here, the index $f$ denotes the neutrino flavor at Earth. Note also that
the response tensor $A_\text{eff}^f(E_\mu, E_\nu)$ (which here has been integrated over
zenith angles) is different for data taken in 2010 and data taken in 2011. Therefore,
we compute the two contributions separately, and then sum them.
Note that $\Delta\Omega= 2\pi$ here since only upward going muons can be reliably
distinguished from background.
With only $2\pi$ of sky coverage, also the numerical prefactor in
\cref{eq:gal-flux} changes. Numerically, we obtain $1.28 \times 10^{-5} \,
\text{cm}^{-2} \, \text{s}^{-1} \, \text{sr}^{-1}$ for the TGM analysis.

For the primary spectrum of the astrophysical component of the neutrino flux,
we consider a simple power law,
\begin{align}
  \frac{dN_{\nu,\text{astro}}}{dE_\nu}
    = J_0 \cdot \bigg( \frac{E_\nu}{100\,\text{TeV}} \bigg)^{-\gamma},
\end{align}
with normalization $J_0$ (in units of $\text{GeV}^{-1}\,\text{cm}^{-2}\,
\text{sec}^{-1}\,\text{sr}^{-1}$) and power law index $\gamma$.
For the astrophysical flux, we assume a flavor ratio of $(1:1:1)$ at Earth,
and we assume equal fluxes of neutrinos and antineutrinos.
The total number of measured events from astrophysical sources is
obtained in analogy to \cref{eq:N-DM,eq:N-DM-TGM}.

\begin{figure}
  \centering
  \begin{tabular}{cc}
    \includegraphics[width=0.48\textwidth]{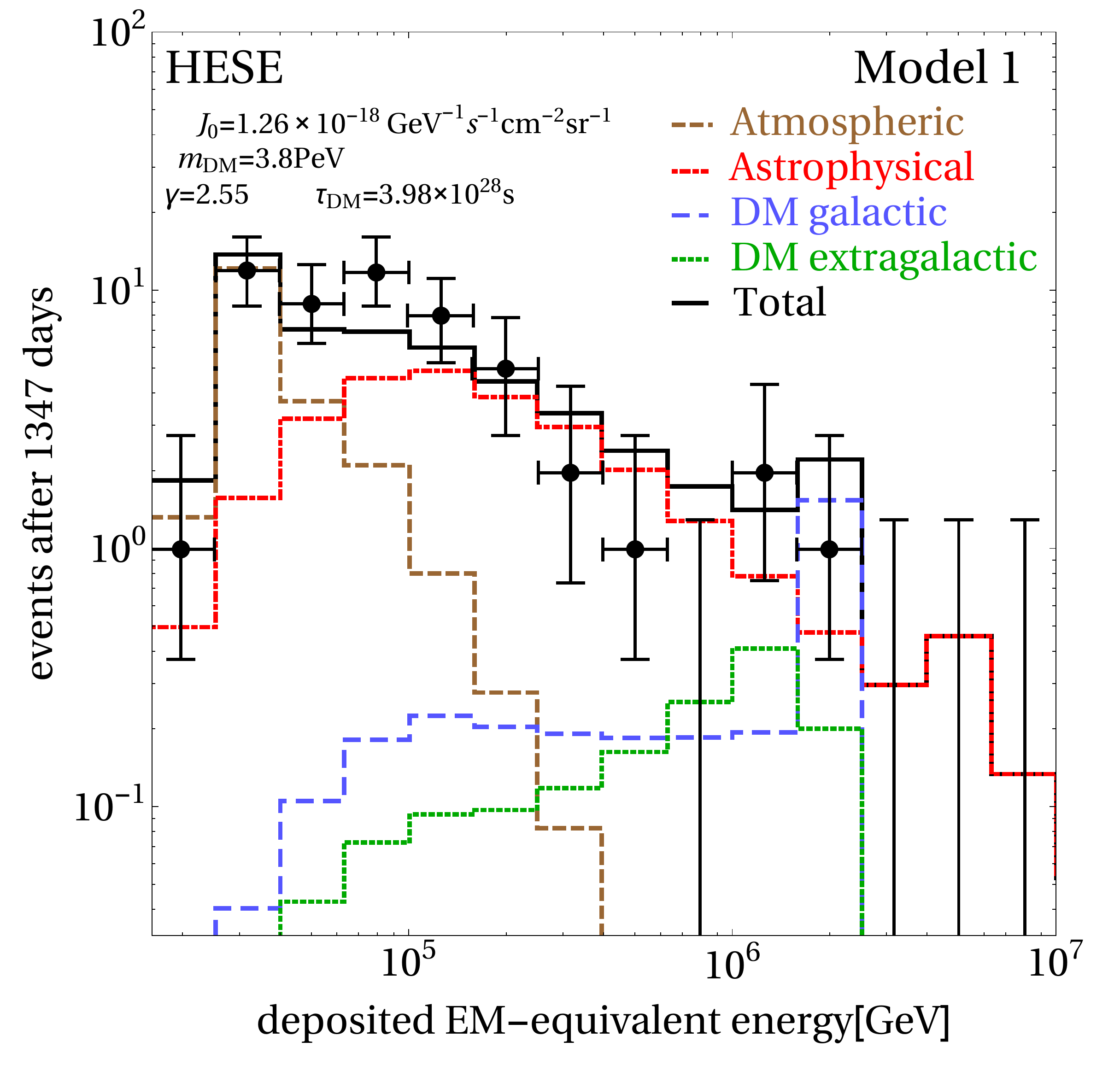} &
    \includegraphics[width=0.48\textwidth]{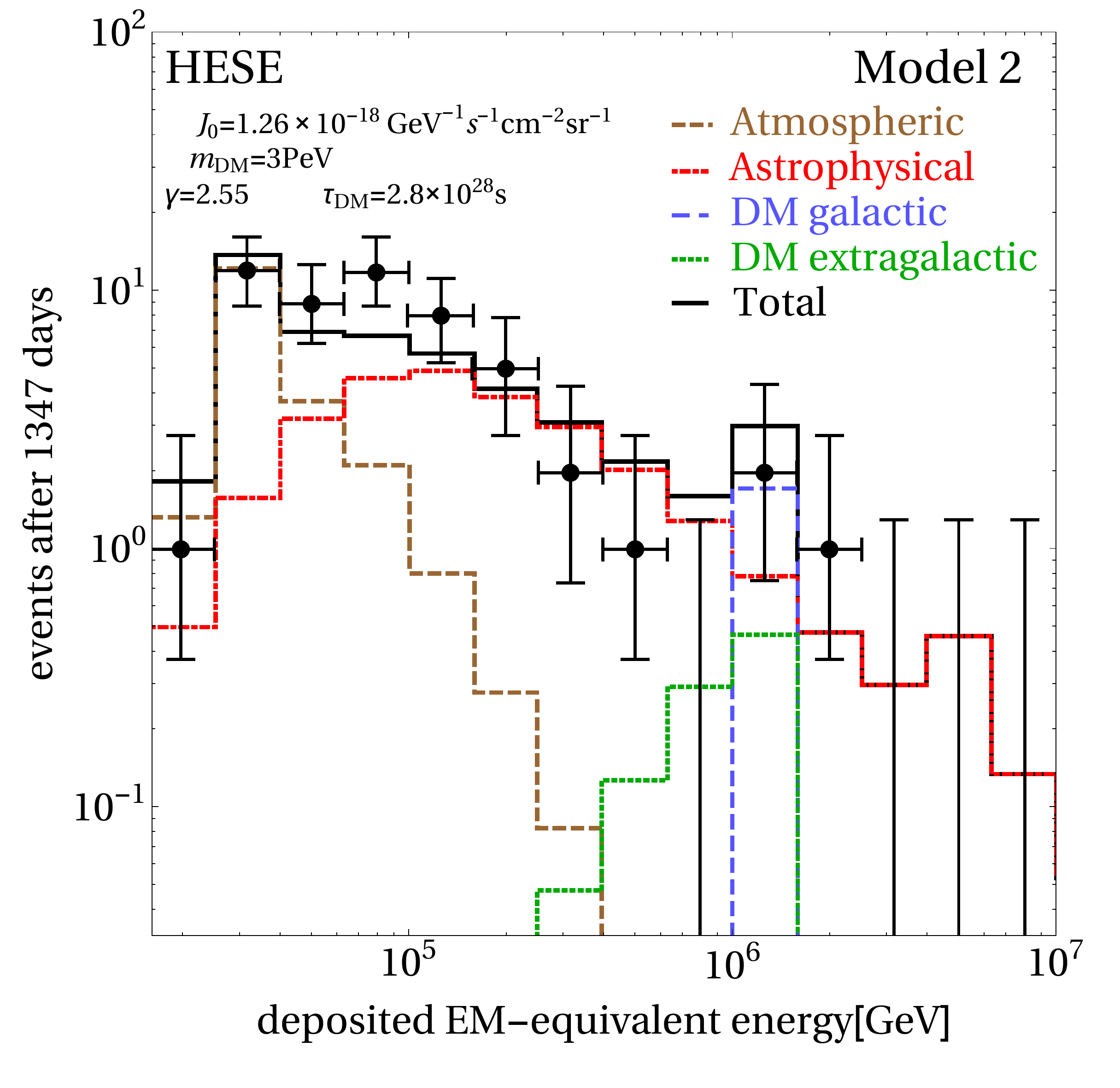} \\
    (a) & (b) \\
    \includegraphics[width=0.48\textwidth]{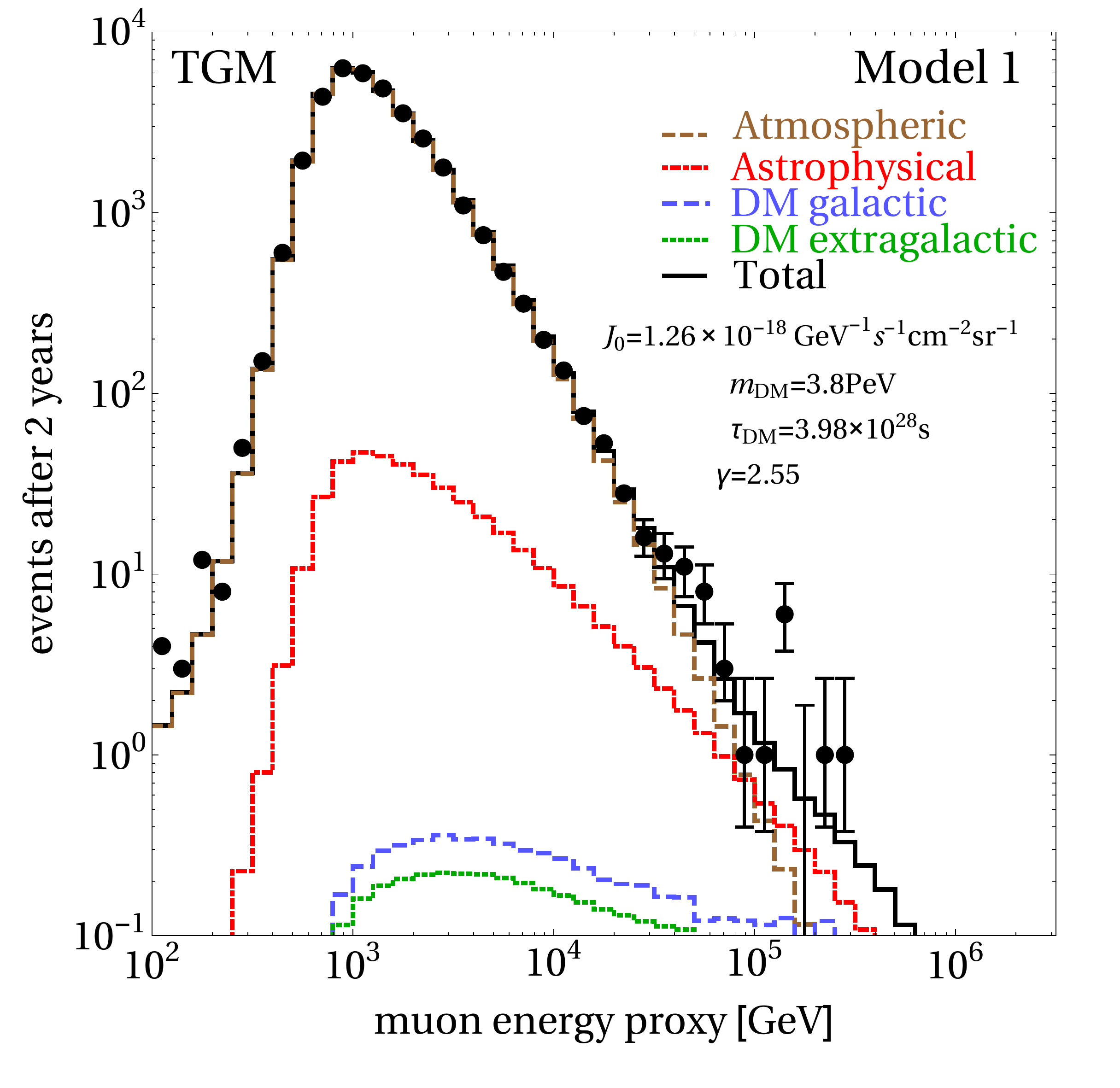} &
    \includegraphics[width=0.48\textwidth]{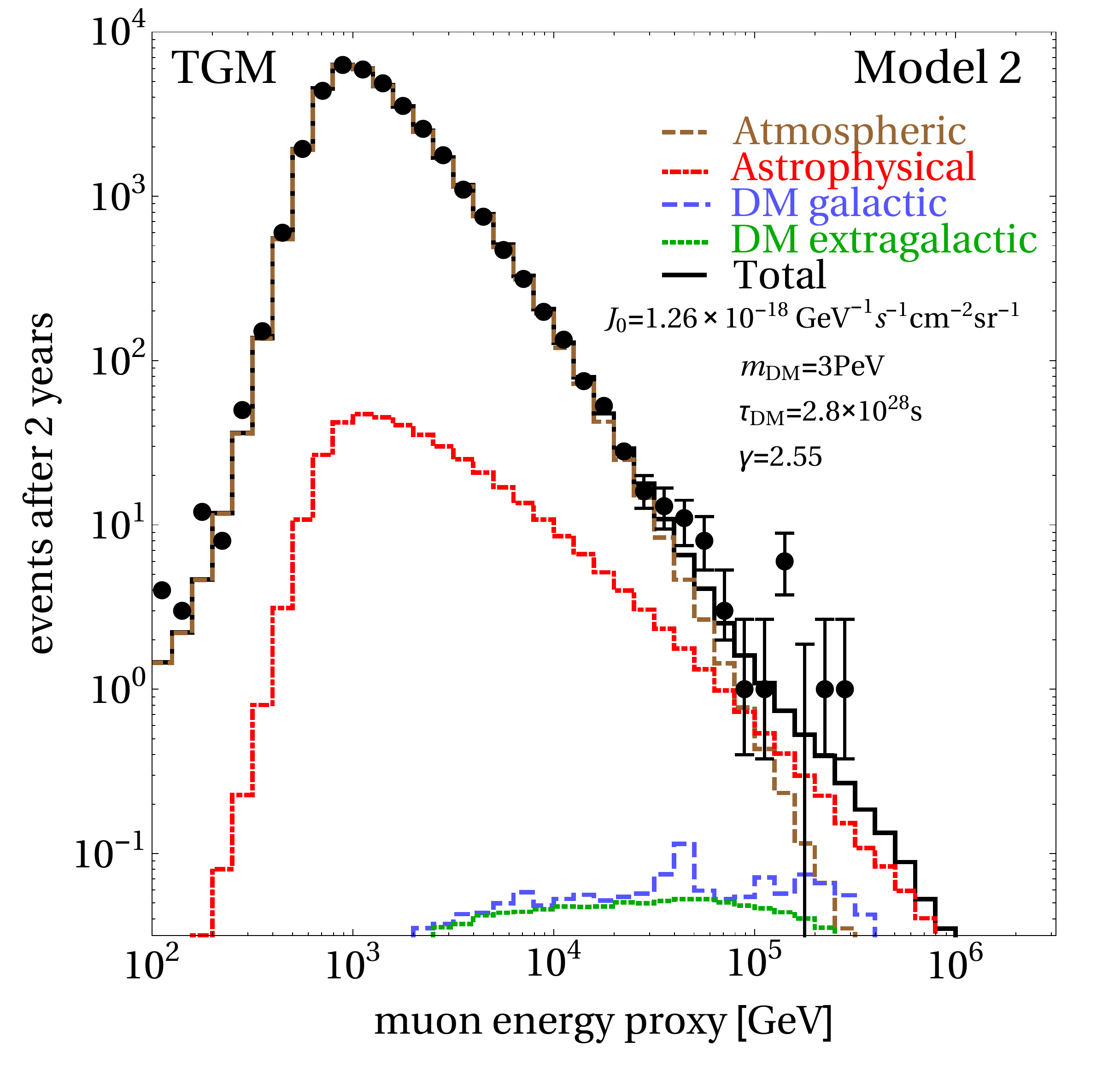} \\
    (c) & (d) \\
  \end{tabular}
  \caption{Predicted neutrino fluxes in model~1 (left) and model~2 (right),
    compared to four-year HESE (high energy starting event) data
    \cite{Aartsen:2015zva, Aartsen:2014gkd} (top)
    and two-year TGM (through-going muon data) \cite{Aartsen:2015rwa}.
    The model parameters (indicated in the plots) are given by the best fit point
    of a combined fit to HESE and TGM data.
  }
  \label{fig:histograms}
\end{figure}

\begin{figure}
  \centering
  \begin{tabular}{cc}
    \includegraphics[width=0.48\textwidth]{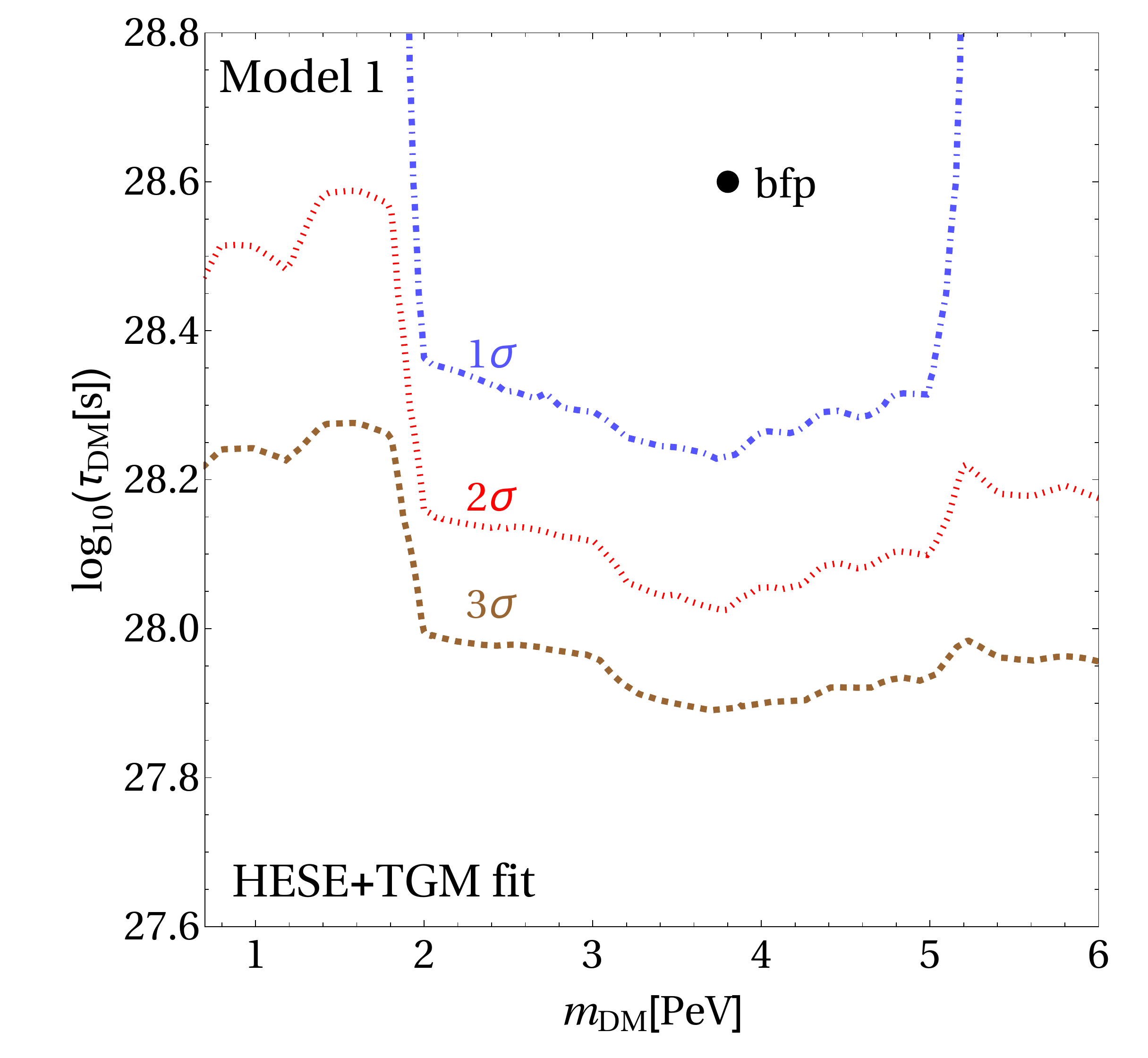} &
    \includegraphics[width=0.48\textwidth]{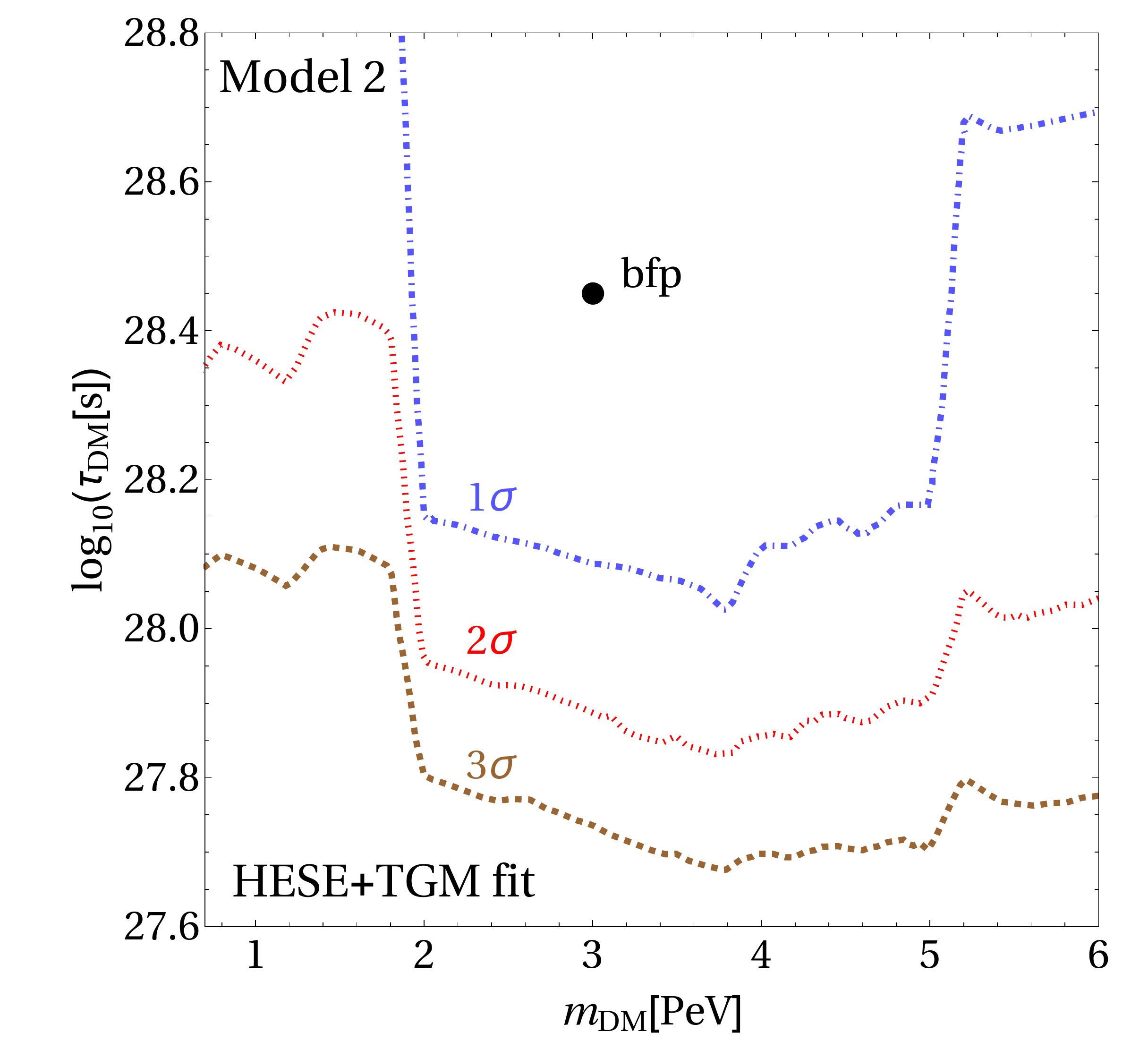} \\
    (a) & (b) \\
    \includegraphics[width=0.48\textwidth]{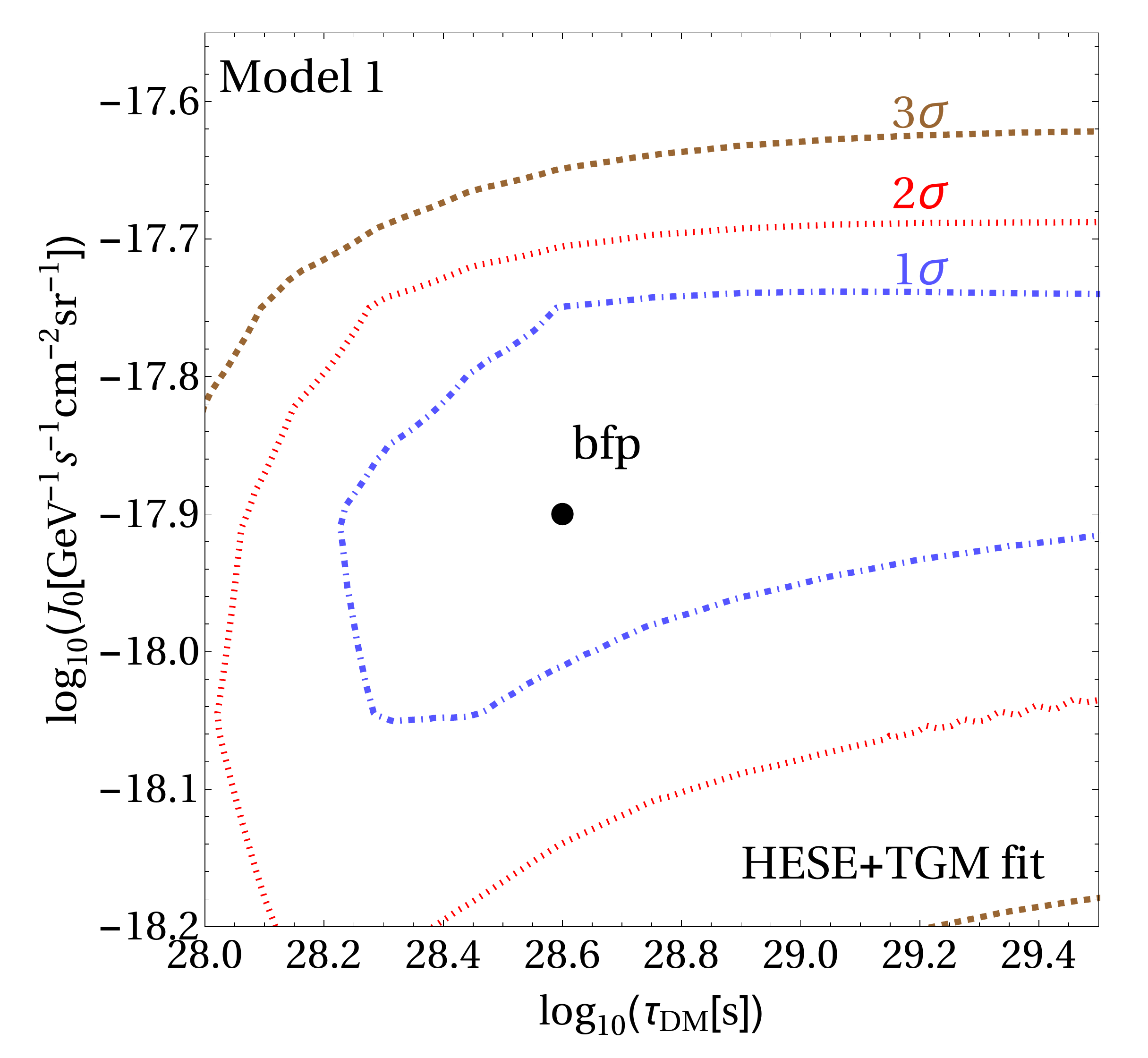} &
    \includegraphics[width=0.48\textwidth]{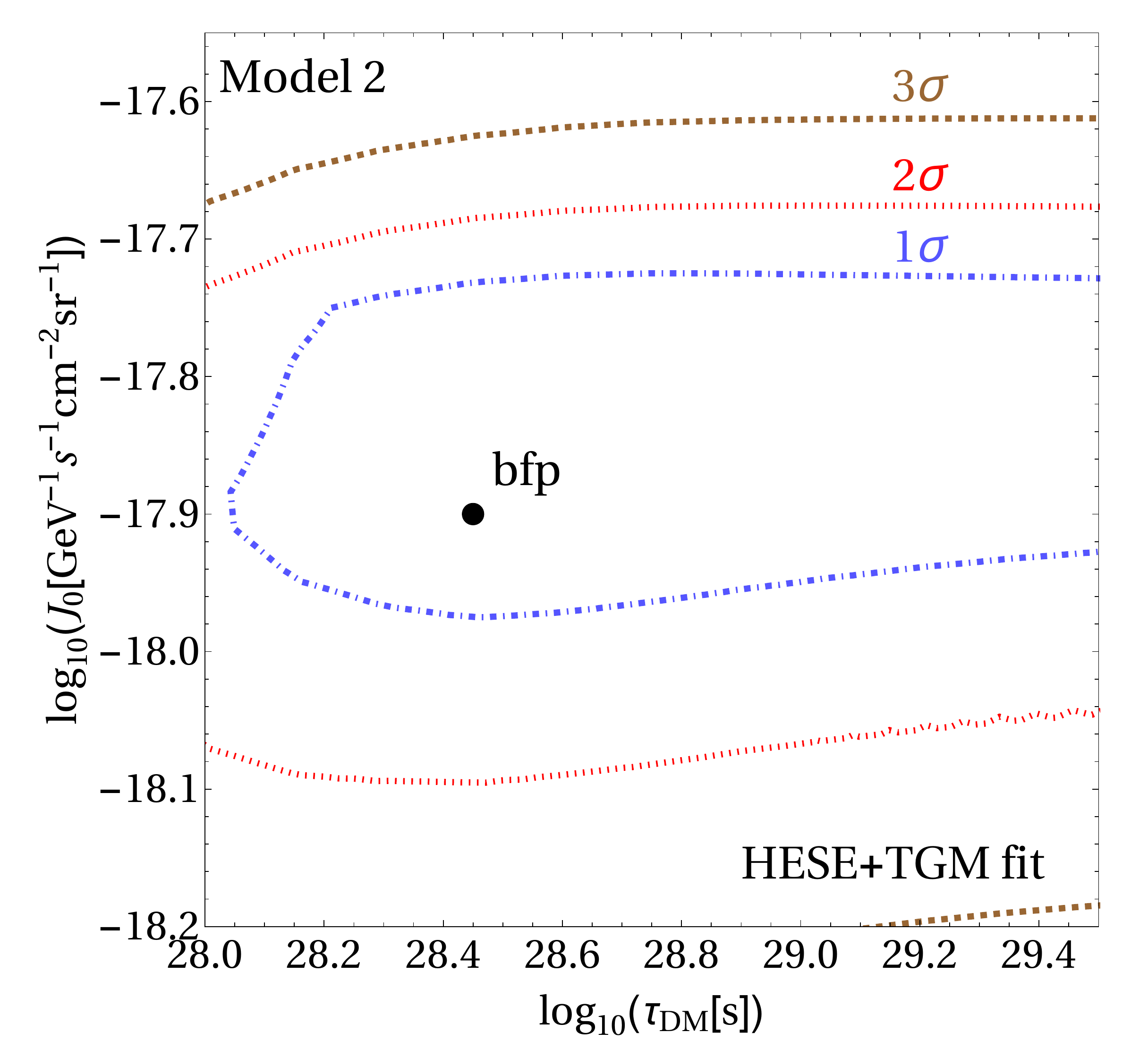} \\
    (c) & (d) \\
  \end{tabular}
  \caption{Results from combined fits to HESE and TGM data. Panels (a) and
    (c) correspond to model~1, while panels (b) and (d) are for model~2.
    In the upper panels, (a) and (b), we show the best fit point (bfp) and
    the $1,2,3\sigma$ preferred regions contours in the plane spanned by
    the dark matter mass $m_\text{DM}$ and its lifetime $\tau_\text{DM}$.
    In the bottom panel, we plot $\tau_\text{DM}$ vs.\ the normalization
    of the astrophysical flux, $J_0$.  In each plot, we project over the
  parameters not shown.} 
  \label{fig:contours}
\end{figure}

We use the log likelihood ratio (LLR) method to determine the parameters in our
toy models.  The LLR is defined as
\begin{align}
&\text{LLR} \bigg(m_{\text{DM}}, \tau_{\text{DM}}, J_0, \gamma \bigg)  \nonumber\\
    &\qquad = \log \left(
        \frac{ \mathop\text{Max}\limits_{x \in [-\infty,\infty]}
               \Big[
                  f_\text{Gauss}(x) \,
                  \prod_i f_\text{Poisson} \Big(
                  S_i \big(m_{\text{DM}}, \tau_{\text{DM}}, J_0, \gamma \big)
                + B_i + x \,\Delta B_i \, \Big| \, O_i \Big) \Big] }
             { \mathop\text{Max}\limits_{x' \in [-\infty,\infty]}
               \Big[
                 f_\text{Gauss}(x') \,
                 \prod_i f_\text{Poisson} \big(
                 B_i + x' \,\Delta B_i \big| O_i \big) \Big] }
      \right) \,.
  \label{eq:chi2-fit}
\end{align}
Here, $f_\text{Gauss}(x)$ is a normal distribution in $x$, with zero mean and
variance 1,
and $f_\text{Poisson}(\mu|n)=\mu^n e^{-n}/n! $ is the Poisson likelihood function. $S_i (m_{\text{DM}}, \tau_{\text{DM}}, J_0, \gamma)$ is the predicted signal event rate
in the $i$-th bin (including astrophysical and DM-induced contributions),
$B_i$ is the atmospheric background, and $\Delta B_i$ is its uncertainty.
Finally, $O_i$ are the observed event rates.
We compute the log-likelihood ratios for both HESE and TGM data and sum them up:
\begin{align}
  \text{LLR}_{\text{total}} \big(m_{\text{DM}}, \tau_{\text{DM}}, J_0, \gamma \big)
    = \text{LLR}_{\text{HESE}} \big(m_{\text{DM}}, \tau_{\text{DM}}, J_0, \gamma \big) +  
      \text{LLR}_{\text{TGM}} \big(m_{\text{DM}}, \tau_{\text{DM}}, J_0, \gamma \big) \,.
\end{align}

The results of our fit are shown in \cref{fig:histograms,fig:contours}.  In
\cref{fig:histograms} we compare the predicted signal and background fluxes at
the respective best fit points for model~1 and model~2 to IceCube data.
Overall, we find an excellent fit, showing that the IceCube data admits (and in
fact prefers) an admixture of neutrinos from DM decay to the astrophysical
neutrino flux.  This is also evident from \cref{fig:contours}, where we explore
the preferred parameter regions in more detail.  We see that the fit, which is
driven by the HESE data, prefers DM masses around a few PeV. In this case, the
HESE events above 1\,PeV could be explained as coming from DM decay, while the
lower energy excess events would be explained by astrophysical sources.
Note that the astrophysical power law index $\gamma$ required to fit the
data is fairly independent of the DM parameters, hence we do not show
it explicitly in \cref{fig:contours}.

Note that the log-likelihood ratio at the best fit point of model~2 is 50, with
the contribution from HESE data being 35 and that from TGM data 15.  A
HESE-only fit yields a LLR of 38 at its best fit point, and a TGM-only fit
yields 19. The fact that the sum of the last two numbers, 57, is larger than
the LLR at the combined best fit, indicates some tension between the TGM and
HESE data sets when interpreted in terms of astrophysical neutrinos plus a
contribution from DM decay. This tension is mostly driven by the astrophysical 
component of the flux, and was found also by the IceCube collaboration in
studies where a neutrino flux from DM decay was not
considered~\cite{Aartsen:2015knd,Aartsen:2016xlq}.

%-----------------------------------------------------------------------------
\section{Summary and Conclusions}
\label{sec:summary}
%-----------------------------------------------------------------------------

In summary, we have explored the possible impact of light sterile neutrinos on
the flavor ratios of astrophysical neutrinos, measured by neutrino telescopes.
We have seen that the accessible regions in the flavor triangle are enlarged,
though not dramatically. The reason are the relatively stringent experimental
constraints on active--sterile neutrino mixing, which we take into account
by running a full global fit to short and long baseline data.

We have then discussed the interesting possibility that the initial flux of
astrophysical neutrinos at production is purely sterile. While this is probably
not true across the energy spectrum, it can happen in a limited energy window,
namely, in scenarios where dark matter decays predominantly to sterile neutrinos
and dominates the neutrino flux in some energy range. In this case, the flavor
ratios at Earth can be very different from those in the 3-flavor case.
The accessible region in the flavor triangle is then shifted towards pure
tau flavor, as constraints on $\nu_\tau$--$\nu_s$ mixing are weakest.

To illustrate what it takes to obtain an unusual initial flavor composition,
we have carried out in the second part of the paper a fit to
IceCube data in two
dark matter toy models: (1) decay of
total singlet DM, $N$, via an operator of the form $\bar{L} H N$, which can
lead to a primary flux dominated by $\nu_\tau$ at $E_\nu \sim m_\text{DM}/2$.
(2) decay of scalar DM to
sterile neutrinos. We have considered high energy starting events (HESE)
and through-going muon (TGM) data. As flavor ratios do not have serious
model discrimination power yet, we have only fitted the energy spectra.
In both toy models, we have found excellent fits to the data,
and we have explored the viable parameter space. This highlights that the
two toy models we consider could be possible targets for future
analyses of IceCube or IceCube Gen-2 data, which could then also take into
account the flavor composition as a function of energy.

%====================================================================%
\section*{Acknowledgments}
%====================================================================%

We would like to thank Carlos Arg\"uelles, Pilar Coloma, Mona Dentler, and Jia
Liu for very useful and interesting discussions.  JK's work was in part carried
out at Fermilab and at the Aspen Center for Physics (NSF grant PHY-1066293),
while VB was visiting the Wisconsin IceCube Particle Astrophysics Center during
the final stages of this project. It is a pleasure to thank these institutions
for their hospitality and support.  This work has been supported by the German
Research Foundation (DFG) under Grant Nos.\ \mbox{KO~4820/1--1}, FOR~2239,
EXC-1098 (PRISMA) and by the European Research Council (ERC) under the European
Union's Horizon 2020 research and innovation programme (grant agreement No.\
637506, ``$\nu$Directions'').

%-----------------------------------------------------------------------------
\bibliographystyle{JHEP}
\bibliography{flavor-ratios}
%-----------------------------------------------------------------------------

\end{document}